# The Cosmic Spacetime

Is The Universe Much Simpler Than we Thought?

## Fulvio Melia

**Cosmology today is confronted with several seemingly insoluble puzzles and strange, inexplicable coincidences. But a careful re-examination of the Cosmological principle and the Weyl postulate, foundational elements in this subject, suggests that we may be missing the point. The observations actually reveal a simpler and more elegant Universe than anyone could have imagined.**

The polish priest Nicolaus Copernicus (1473—1543) started a revolution with his heliocentric cosmology that displaced the Earth from the center of the Universe. His remarkable shift in paradigm continues to this day, the cornerstone of a concept we now call the Cosmological Principle, in which the Universe is assumed to be homogeneous and isotropic, without a center or boundary. But few realize that even this high degree of symmetry is insufficient for cosmologists to build a practical model of the Universe from the equations of General Relativity.

The missing ingredient emerged from the work of mathematician Hermann Weyl (1885—1955), who reasoned that on large scales the Universe must be expanding in an orderly fashion. He argued that all galaxies move away from each other, except for the odd collision or two due to some peculiar motion on top of the ``Hubble flow'' (figure 1). In this view, the evolution of the universe is a time-ordered sequence of 3-dimensional space-like hypersurfaces, each of which satisfies the Cosmological Principle—an intuitive picture of regularity formally expressed as the *Weyl postulate*.

Together, these two philosophical inputs allow us to use a special time coordinate, called the *cosmic time* $t$, to represent how much change has occurred since the big bang, irrespective of location. In special relativity, this approach can be confusing because $t$ is the proper time on a clock at rest with respect to the observer, but is not the time she would measure on her synchronized clocks at other locations. But since the physical conditions are presumably the same everywhere, $t$ should track the evolution of the Universe as seen from any vantage point, since the same degree of change will have occurred anywhere on a given time slice shown in figure 1.

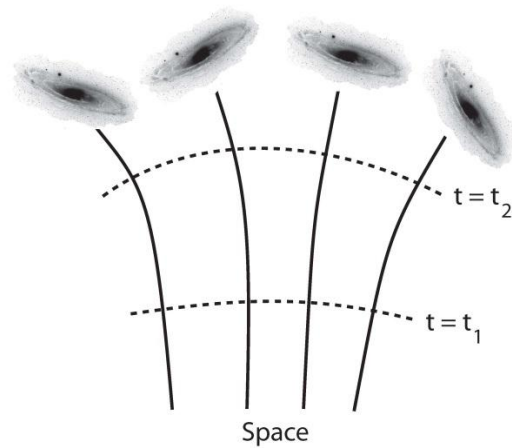

**Figure 1.** Illustration of the Weyl postulate.

Of course, the Weyl postulate has several other important consequences, particularly in terms of how we interpret the separation between any two points in the cosmic flow. In relativity, the *proper distance* $R(t)$ between two points is their separation measured simultaneously in a given frame. It is not difficult to convince oneself that if the distance $R_{AC}$ between frames A and C (figure 2) is twice $R_{AB}$, then A and C must be receding from each other at twice the rate of A and B. The Weyl postulate therefore reduces formally to the mathematical expression

$R(t) = a(t)r$, meaning that the proper distance $R(t)$ between any two spacetime points must be the product of a fixed, co-moving distance $r$—which never changes even as the Universe expands—and a universal function of time $a(t)$ independent of position, but not necessarily of time.

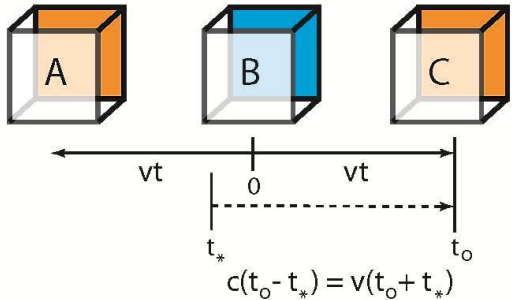

**Figure 2**. A and C recede from us (in B) at the same speed v. At time $t_*$, a light pulse is emitted by A that arrives at C at time $t_0$. All times in this diagram are measured on our clock.

A galaxy a proper distance $R(t)$ from us must therefore be receding at speed $\dot{R} = \dot{a}r$, or $\dot{R} = (\dot{a}/a)ar$, conventionally written $\dot{R} = HR$ (the "Hubble Law"). The Hubble constant $H \equiv \dot{a}/a$ is independent of position. Notice how easily such a simple consequence of the Weyl postulate accounts for all of Hubble's famous observations, which convinced even Einstein—an early advocate of the static Universe—that the cosmos is expanding at a speed proportional to proper distance. There is much to learn about the Universe—perhaps even its beginnings—by understanding $H$ or $a(t)$, and so much of the effort in cosmological research is dedicated to this task.

### The Earliest Moments

But when we attempt to follow what happened in the earliest moments, we immediately hit a roadblock because General Relativity is a theory of precision, whereas quantum mechanics imposes an irreducible fuzziness on any measurement of distance or time. Quantum physicists argue that the precision of particle location is no better than its Compton wavelength, $\lambda_C$, the wavelength that a photon would have if its energy were equal to that of the particle with rest mass $m$ (i.e., $mc^2$). This makes sense because to locate the particle, you would want to shine the highest frequency light on it, except that if the photon's energy is too high, the particle recoils and you again lose track of where it is. The ideal compromise is realized when their energies are equal, so $\lambda_C = h/mc$.

But what value of $m$ should one use to calculate $\lambda_C$? Although we don't yet have a theory of quantum gravity, it is nonetheless reasonable to suppose that such a unification would exhibit its strongest effects in the early universe. Thus, instead of using any particular particle's mass to define the Compton wavelength, cosmologists equate it to a length scale from relativity, the argument being that there must have been a mutual consistency between the various physical factors that define each theory. So they use the radius of curvature associated with a mass so compact that it wraps itself with an event horizon. This radius, $R_h$, formally derived by Karl Schwarzschild (1873—1916), was actually known classically as the radius a mass would need to have in order for the escape speed at its surface to equal the speed of light, so $c^2/2 = Gm/R_h$, or $R_h = 2Gm/c^2$.

Equating the Compton wavelength to the Schwarzschild radius yields a unique value for $m$, i.e., $m_P = 5 \times 10^{-8}$ kg, known as the Planck mass. And the Compton wavelength for this mass, known as the Planck length, is therefore $l_P = \lambda_C(m_P) = 10^{-33}$ m, roughly $10^{-20}$ times the radius of the proton. This is believed to be the smallest distance about which anything can be known. We can also estimate the shortest time interval associated with the Planck scale, essentially the light-crossing time of a Planck length, known as the Planck time, $t_P = l_P/c \approx 10^{-43}$ seconds.

These physical scales constitute the starting point for any discussion of the cosmic spacetime. But notice their dependence on the Schwarzschild radius, which *delimits* the volume of interest. Today we know that the Universe is infinite, so it must always have been infinite, even at the beginning. What relevance, then, can the Schwarzschild radius have to the cosmic spacetime?

### The Universe's Gravitational Radius

The answer may be found in a theorem published in 1923 by George David Birkhoff (1884—1944), a preeminent mathematician at Harvard University [1]. Our intuition tells us that because of the Cosmological Principle, an observer must experience zero *net* acceleration from a mass distributed isotropically around him. But in fact the *relative* acceleration between an observer and any other point in the cosmos is not zero; it depends on the mass-energy content between himself and that other point. The Birkhoff theorem, and its corollary, can help us understand the difference between these two viewpoints.

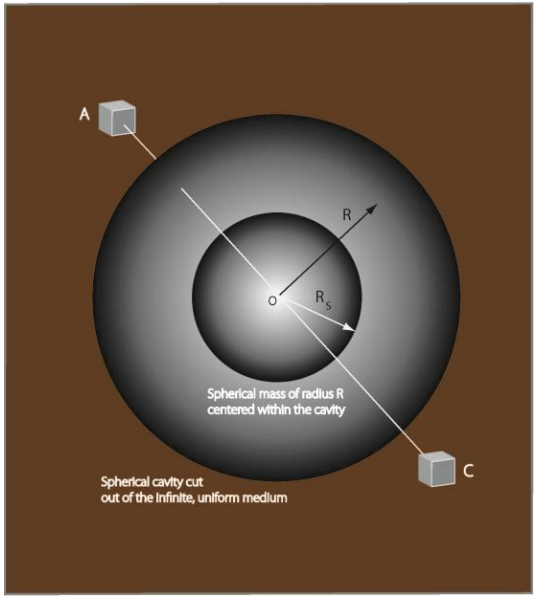

**Figure 3**. Spherical mass of proper radius $R_S$ inside a spherical cavity cut out of an otherwise uniform, infinite medium. A and C are two small parcels of mass-energy contributing to the gravitational field inside the cavity.

Birkhoff's theorem is a relativistic generalization of Sir Isaac Newton's (1642—1727) theorem—that the gravitational field outside of a spherically symmetric body is indistinguishable from that of the same mass concentrated at its center. The corollary to this theorem states that the metric inside a spherical cavity (figure 3) cut out of a uniform medium is equivalent to the flat-space Minkowski metric (i.e., a spacetime without any curvature induced by mass-energy). A simple heuristic argument for this result is based on the fact that for every parcel of mass-energy A outside the cavity (see figure 3), there exists an equal, but opposite, parcel C that—due to the symmetry—completely cancels the effect of the former.

To understand the emergence of a gravitational radius in cosmology, let us now imagine placing an observer at the center of this spherical cavity (whose proper radius is $R_{cav}$), and then surrounding her with a spherically-symmetric mass with a surface of proper radius $R_s < R_{cav}$. The metric in the space between the mass and the edge of the cavity is given by the Schwarzschild solution describing the spacetime surrounding a compact mass, and the relative acceleration between the observer and $R_s$ is simply due to the mass, $M(R_s) = V \rho/c^2$, enclosed within this radius, where $V = (4\pi/3)R_s^3$ is the proper volume. If we keep increasing $R_s$ while keeping $\rho$ constant, we eventually reach a threshold of enclosed mass for which $R_S$ becomes the gravitational horizon $R_h = \sqrt{(3c^2/4\pi)\rho}$.

Observational cosmology is now in a position to actually measure this radius, but before considering this, let us first examine several other aspects of the cosmic spacetime, including a rather surprising connection between $R_h$ and the Hubble law.

### Dynamics

To fully account for the expansion of the Universe in terms of its constituents we must introduce dynamical equations. We can arrive at the most important of these using the Birkhoff theorem in the Newtonian limit. Known as the Friedmann equation, after Alexander Friedmann (1888—1925), this equation is an expression of the (classical) conservation of energy.

Imagine placing a particle of mass $m$ on the surface of the spherical mass in figure 3. Because of the Birkhoff theorem, the behavior of this particle relative to the observer at the origin is dictated solely by the mass-energy contained within $R_s$. Classically, the particle's energy is

$$E = \tfrac{1}{2} m \dot{R}_s^2 - \frac{GM(R_s)m}{R_s}.$$

Replacing $R_s$ with its Weyl form, $R_s = a(t) r_s$, one gets

$$H^2 = \left(\frac{\dot{a}}{a}\right)^2 = \frac{8\pi G}{3c^2}\rho - \frac{kc^2}{a^2},$$

where $k$ is a constant proportional to the total energy, $E$, of the particle. This equation is identical to that formally derived from General Relativity.

The cosmological observations [2] seem to indicate that the Universe is flat, meaning that $k = 0$. However, based on what we now know about $k$, we conclude that the total energy in the Universe must be exactly zero, since every particle in the cosmos will satisfy the Friedmann equation. The big bang somehow separated positive kinetic energy from negative potential energy, but balanced precisely, lending support to the notion that the Universe began as a quantum fluctuation in vacuum.

Setting $k=0$, it is easy to see that $R_h = c/H$, another easily recognizable (and observationally important) quantity known as the "Hubble" radius. Notice in the Hubble law, $\dot{R} = HR$, that the speed of recession reaches $c$ at the radius $R = c/H$, which we now understand is simply the Universe's gravitational horizon, $R_h$. Physicists who study black holes already know that an event horizon approaches a free-falling observer at speed $c$. And now we recognize the same phenomenon occurring in cosmology, since $\dot{R} = c$ at the Universe's own gravitational horizon, $R_h$.

But here is perhaps the most telling indicator of what the Universe may be doing. Since we now know that $R_h = c/H$, we can determine the Universe's current gravitational radius from the measured Hubble constant $H \approx 70$ km/s/Mpc [3]. Thus, $R_h \approx 13.3$ billion light years, oddly (one should say, amazingly) close to the estimated maximum distance, $ct_0$, light could have traveled during the time ($t_0$) since the big bang, because in the standard model of cosmology (which we will introduce shortly), the estimated age of the Universe is $t_0 \approx 13.7$ billion years. The gravitational radius $R_h$ could have been anything; its apparent equality to $ct_0$ must be telling us something that we should not ignore. We shall see shortly that the identification of the Hubble radius as the Universe's gravitational horizon unavoidably implies a rather profound conclusion concerning the Universal expansion.

## The $R_h = ct$ Universe

The Weyl postulate compels us to treat every proper distance as the product of the universal expansion factor $a(t)$ and an unchanging co-moving distance $r$. Therefore, since we defined $R_h$ in terms of the proper volume enclosing the mass $M(R_h)$, the gravitational radius must itself be a proper distance, $R_h = a(t) r_h$, where $r_h$ remains constant as the Universe expands. But then $R_h = c(a/\dot{a})$, and so $\dot{a}$ itself must be a constant in time. The Weyl postulate thus implies that $a(t) \propto t$, which in turn means that $R_h = ct$ [4].

Could the Universe really be this simple? Well, let's look at the evidence. The standard model of cosmology, known as ΛCDM, is a specific choice of $\rho$, comprising matter (visible and dark), radiation, and an unknown "dark energy." Each of these ingredients changes with $a(t)$ in a different way, so it is not easy to obtain a simple analytic solution for the expansion factor. What is known, however, is that radiation probably dominated early on, whereas matter and—more recently—dark energy, seem to be dominating today. And since $\rho \sim a^{-4}$ for radiation, while $\rho \sim a^{-3}$ for matter, and $\rho \sim$ constant if dark matter were a cosmological constant $\Lambda$, one can easily show that $a(t) \sim t^{1/2}$ early on, transitioning to $t^{2/3}$ in the matter-dominated era, and finally becoming an exponential in the future.

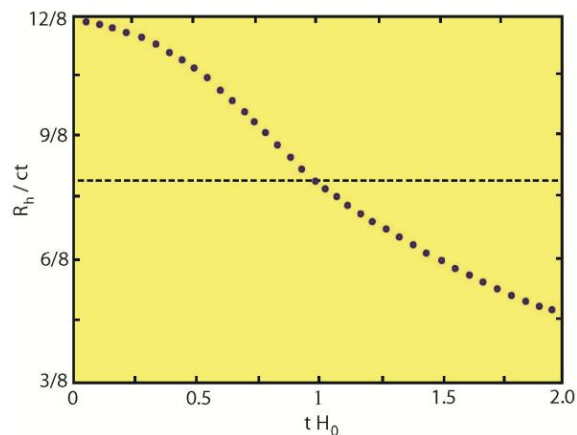

**Figure 4**. The Universe's gravitational radius $R_h$, in units of the distance light could have traveled during a time $t$, as a function of cosmic time, in units of $H_0^{-1}$.

The standard model of cosmology therefore predicts quite a complex pattern of behavior—and yet, through it all, the Universe appears to have undergone just the right amount of deceleration at the beginning, followed by just a precise amount of acceleration in recent times, in order for it to be left with an observed value of $R_h$ equal to $ct_0$ today, which is what we would have gotten anyway if the Universe had simply expanded at a constant rate all along. One may chalk this up to mere coincidence, but closer scrutiny shows that this is simply untenable. One can easily calculate the ratio $R_h/ct$ as a function of time (figure 4), since we know how $\rho$ changes with $a(t)$. The horizontal dashed line corresponds to the observed value $R_h(t_0)/ct_0 = 1$. Clearly, within the context of ΛCDM, this special condition can be met only once in the entire history of the Universe, *right now*, when we just happen to be looking. Really?

Though one cannot completely rule out such an amazing coincidence, it is far more satisfying philosophically to interpret these observations as meaning that, though ΛCDM may be a workable approximation to the "real" Universe in some instances, the actual behavior of the Universe is more in line with the Weyl postulate, which requires that $R_h$ be always equal to $ct$, and $a(t) \propto t$. This would then explain why we see $R_h(t_0) = ct_0$ today (because these two distances are *always* equal to each other), and it would explain why the physical conditions in the early Universe are consistent with the use of the Schwarzschild radius to balance the Compton wavelength.

## Inflation

Several other arguments in favor of the $R_h = ct$ Universe are documented in the references cited at the end of this article. One particular consequence of the $R_h = ct$ condition stands out because it would immediately obviate a major, long-standing problem with the standard model, having to do with cosmology's 30-year (unsuccessful) odyssey with inflation.

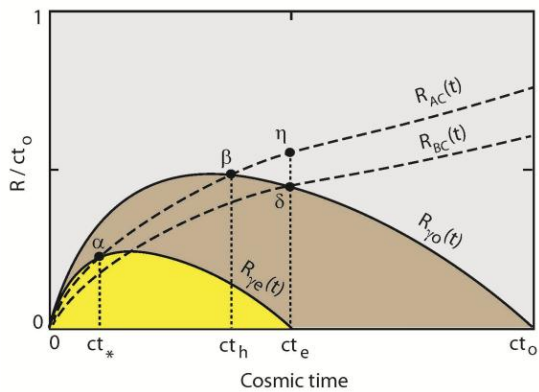

**Figure 5**. Proper distance for light, $R_{\gamma 0}$ and $R_{\gamma e}$, approaching the observer C (in figure 2) at times $t_0$ and $t_e$, respectively, as a function of cosmic time, together with the proper distances (relative to C) of sources A and B as functions of this same $t$.

The so-called horizon problem in cosmology is viewed as a major shortcoming of the standard model because the Universe seems to have required special initial conditions that are highly improbable. The horizon problem arises from the observed uniformity of the microwave background radiation, which has the same temperature everywhere, save for fluctuations at the level of one part in 100,000. Regions on opposite sides of the sky, the argument goes, lie beyond each other's horizon, yet their present temperature is identical, even though they could not possibly have ever been in thermal equilibrium.

This deficiency of the standard model is best understood with the diagram in figure 5. Within the context of ΛCDM, one can solve for $a(t)$ and determine the trajectories of light reaching the observer. With reference to elements A, B, and C in figure 2, this diagram shows which sources (A or B) could have emitted light at specific times (corresponding to, say, points $\alpha$ and $\delta$) that reaches C at either time $t_e$ or $t_0$. From our perspective in B, we see light emitted by A and C at time $t_e$, when they were apparently in equilibrium. Therefore, light emitted by one of them, say A at time $t_*$, must have reached the other by the time ($t_e$) they produced the cosmic microwave background we see today at time $t_0$. But in ΛCDM, the deceleration that occurred following the big bang would have made it impossible for $R_{AB}$ to have been equal to $R_{BC}$ at $t_e$.

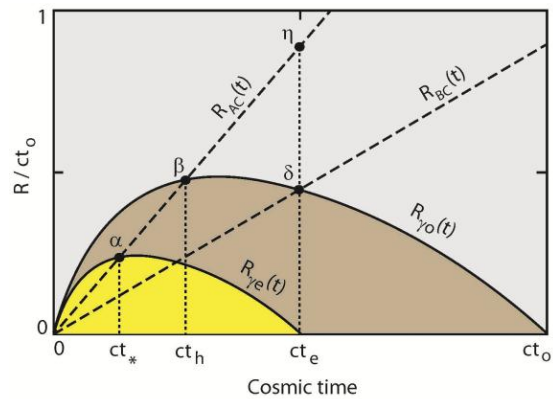

**Figure 6**. Same as figure 5, except for the $R_h = ct$ Universe. Note that in this case, $R_{AC}(t_e) = 2R_{BC}(t_e)$, which means that A and C (in figure 2) were equidistant from us (in B) at the time they emitted the light we see today as the cosmic microwave background radiation.

So serious has this shortcoming become that the inflationary model of cosmology [5] was invented to resolve this possible discrepancy. In this picture, an inflationary spurt occurred at $10^{-35}$ seconds following the big bang, carrying causally connected regions beyond the horizon each would have had in the absence of this temporary acceleration.

But after three decades of struggling with this "fix," it is now clear that inflation may not be the solution after all. The idea of inflation is itself fraught with many apparently insurmountable problems. For example, monopoles should have been produced copiously in Grand Unified Theories at high temperature in the early Universe, and should be so prevalent today that they should be the primary constituent of the Universe. Yet they have never been found.

The $R_h = ct$ Universe easily resolves this issue because it completely does away with the so-called horizon problem [6]. This simple Universe therefore does not need inflation to have occurred. To understand why, we need to consider the diagram shown in figure 6. The principal difference between the two cases is that, here, there was no early deceleration in the Universe, and therefore it was possible for $R_{AB}(t_e) = R_{BC}(t_e)$, while still maintaining causal contact between A and C before they emitted the light we see today.

## Conclusion

The observed equality between $R_h$ and $ct$ today may be hinting at a great simplification to our view of the Universe. In fact, the Cosmological Principle and Weyl's postulate together require a constant expansion rate $a(t) \propto t$. With this simplification, we inherit several significant improvements to our understanding of how the Universe began and how it has evolved to this day. We understand why the Schwarzschild radius was relevant to the early Universe (because it is always relevant), we understand why $R_h = ct$ today (because they are always equal), and we can understand how the Universe could have functioned without inflation, an idea that has never quite solidified into a fully consistent theory. How simple. Who would have thought?

## About the Author

Fulvio Melia is Professor of Physics, Astronomy, and the Applied Math Program at the University of Arizona, in Tucson, and John Woodruff Simpson Fellow at Amherst College, a chair formerly held by the Nobel laureate Neils Bohr and noted American Poet Robert Frost. Born in Gorizia, Italy, he was raised in Melbourne and received his BSc and MS degrees from Melbourne University. He completed his graduate studies at the Massachusetts Institute of Technology, receiving a PhD for research on the physics of strong gravitational and magnetic fields. Since then, he has been a Presidential Young Investigator (under Ronald Reagan) and an Alfred P. Sloan Research Fellow. He has been a visiting professor at numerous universities, both here in Australasia and in Europe, and has published over 250 journal articles in high-energy astrophysics, including topics on black holes, relativistic matter, and cosmology. He is also the author of 6 books, most recently *Cracking the Einstein Code,* the story of how New Zealander Roy Kerr and his colleagues finally managed to solve Einstein's equations of General Relativity.

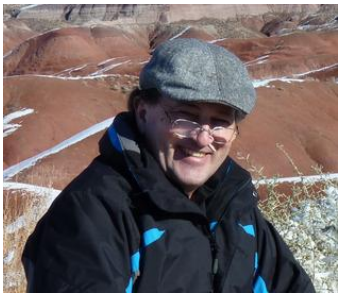